\documentclass[%
 reprint,
 amsmath,amssymb,
 aps,
]{revtex4-1}

\usepackage{graphicx}
\usepackage{dcolumn}
\usepackage{bm}
\usepackage{graphicx}
\usepackage{dcolumn}
\usepackage{bm}
\usepackage{bigints}
\usepackage{color}
\usepackage{amsmath,amssymb,graphicx}
\usepackage{float}
\usepackage{braket}
\usepackage{amsmath}
\usepackage{textcomp}

\begin{document}

\newenvironment{packedit1}{
\begin{itemize}
  \setlength{\itemsep}{1pt}
  \setlength{\parskip}{0pt}
  \setlength{\parsep}{0pt}
}{\end{itemize}}

\preprint{APS/123-QED}

\title{Temperature distribution inside a double-cladding optical fiber laser or amplifier}

\author{Arash Mafi}
\affiliation{Department of Physics \& Astronomy and Center for High Technology Materials, University of New Mexico, Albuquerque, New Mexico 87131, USA\\
             mafi@unm.edu}%

\date{\today}

\begin{abstract}
The temperature distribution inside a double-cladding optical fiber laser or amplifier is examined in detail. Traditionally, the quantum defect in the core is taken to be the main source of heating in an active optical fiber. However,  contributions from the parasitic absorption of the signal and the pump may also play an important role, especially for low quantum defect or radiation-balanced lasers and amplifiers. The contributions to the heating in both the core and the inner-cladding are considered and analyzed in general terms in this paper. In particular, it is shown that if the maximum tolerable surface temperature of the fiber relative to the ambient is taken to be 300 degrees Celsius to avoid damaging the fiber's outer polymer cladding, the core temperature rises only in the range of 0-5 degrees Celsius relative to the inner-cladding for an air-cooled fiber. However, for a water-cooled fiber, the core temperature can be higher than the inner-cladding by as much as 50 degrees Celsius, potentially changing a single-mode core to multimode due to the thermo-optic effect.
\end{abstract}


\maketitle
\section{Introduction}
\label{eq:intro}
The power generated from optical fiber lasers and amplifiers has increased significantly over the past decade~\cite{zenteno1993high,richardson2010high,zervas2014high,Tunnermann}. 
Consequently, efficient heat mitigation has become one of the main concerns, especially in light of recent reports of limitations in power scaling because of the thermally-induced 
mode instability, which degrades the output beam quality~\cite{Smith,ward2012origin,jauregui1,jauregui2012physical,Scarnera}. There already exists a sizable body of literature
on the thermal analysis of optical fiber lasers and amplifiers~\cite{Brown,LiLi,Hadrich,Fan,Hansen,Mousavi1,Mousavi2}. In particular, Brown and Hoffman in Ref.~\cite{Brown} derived 
detailed analytical equations for the temperature distribution inside an optical fiber, assuming that the heat is generated only within the the fiber core. This assumption
is usually valid when the primary source of heating is the quantum defect in the core of a double-cladding fiber (DCF) design; the quantum defect 
being the energy difference between the pump and signal photons. However, in some modern high-power fiber lasers and amplifiers, 
where the quantum defect is lowered~\cite{Li}, or when the amplifier operates in a nearly radiation-balanced regime or for radiation-balanced lasers~\cite{bowman2010minimizing,bowman2016low,mobini2018thermal,mobiniCoreCladding}, the heat generated due to the parasitic absorption of the high-power pump in the inner-cladding
can be considerable and must be included in the analysis. In this paper, we derive analytical expressions for the temperature distribution inside a
DCF laser or amplifier, for a more general case where the heating occurs both in the core and the inner-cladding, albeit at different rates. 
We show that the temperature distributions can be easily calculated for various scenarios. In essence, a single parameter $\gamma$ allows one to interpolate between
the case where the quantum defect heating is dominant to when the parasitic absorption heating is comparable in size or is even the dominant source of heating.
The results can be readily applied to the single-cladding design of a core-pumped fiber as a special case. 

\begin{figure}[t]
\centering
\includegraphics[width=1.7in]{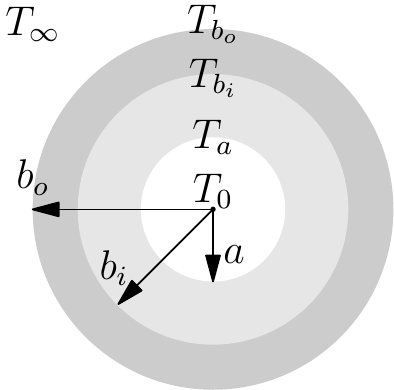}
\caption{Schematic of a DCF with temperature markings.}
\label{fig:fiber-temp}
\end{figure}
A schematic transverse profile of the DCF is shown in Fig.\ref{fig:fiber-temp}, where we assume a cylindrical geometry for the fiber. The core, in which the signal propagates, is marked with the inner white-filled circle of radius $a$ and is doped with rare-earth ions (typically Yb).
The inner-cladding, in which the pump propagates, is the region marked with the light-gray region of radius $b$. Of course, some of the propagating pump power overlaps the 
core region, which is responsible for pumping the core. The outer-cladding of the fiber is the region shaded in dark-gray with radius $c$,
where $D=2c$ is the total outer diameter of the fiber. We also mark the temperature of the center of the fiber core as $T_0$, at the core-inner-cladding boundary as $T_{a}$, 
at the inner-outer-cladding boundary as $T_b$, and the outer surface of the fiber as $T_c$. The ambient outside temperature is identified as $T_\infty$.

Before, we start our analysis, we would like to present a key result obtained in this paper:
\begin{align}
\label{Eq:1stresult}
\delta T_a=\mathfrak{X}_a\dfrac{D}{\mathcal{D}_a}\Delta T.
\end{align}
Here, $\delta T_a=T_0-T_{a}$ is the temperature variation inside the core of the optical fiber, and $\Delta T=T_{c}-T_\infty$ is the 
difference between the surface temperature of the fiber and the ambient temperature. 
We also have:
\begin{align}
\label{Eq:Dth}
\mathcal{D}_a=\dfrac{4\kappa_a}{H},
\end{align}
where $\kappa_a$ is the thermal conductivity of the (fused silica) glass in units of ${\rm W/(m.K)}$ and $H$ is the convective heat transfer coefficient 
in units of ${\rm W/(m^{2}.k)}$ (typically that of air or water). $\mathfrak{X}_a$ is an order one coefficient that is to be determined and depends on 
the geometrical and optical properties of the laser or amplifier.
We note that the fiber surface cannot be feasibly hotter than a few hundred degrees Celsius, so $\Delta T \lesssim 300^{\circ}$C is generally assumed in this paper 
unless stated otherwise.
$D/\mathcal{D}_a$ is proportional to the Biot number of the thermal problem, which is a dimensionless 
quantity used in heat transfer calculations~\cite{book}. The ratio $D/\mathcal{D}_a$ sets the scale for $\delta T_a$ for a given $\Delta T$--for 
air-cooling, $\mathcal{D}_a$ is nearly two orders of magnitude larger than $D$, while it is only an order of magnitude larger for water-cooling.
As such, in water-cooled systems, the core temperature can increase to the point that the fiber changes from supporting only a single-mode to supporting 
multiple modes due to the thermo-optic effect. However, for air-cooled DCFs with $\Delta T\lesssim 300^{\circ}$C, the core temperature rises 
only in the range of $0-5^{\circ}$C relative to the inner-cladding, so the thermo-optic effect is less pronounced and single-mode to multimode
transition may not happen.

In the following, we will present the problem in the most general terms, while providing specific examples to illuminate the main points. 
In Section~\ref{sec:formulation}, we will formulate the problem and derive the relevant equations and results. In Section~\ref{sec:examples}, we will apply our formalism to 
a few specific examples of fibers commonly used in laser and amplifier systems and examine our general conclusions using specific 
numerical examples. In Section~\ref{sec:summary}, we will summarize and conclude. Appendices A and B provide further information about the assumptions used
in deriving the analytical expressions. In Appendix C, we summarize the equations that are most useful for direct comparison with experiments.
\section{Formulation}
\label{sec:formulation}
In this paper, we will refer to the pump laser as the ``pump'', and to the generated laser (in a laser design) or the amplified laser (in an amplifier design) as the ``signal''.
We present our arguments and observations in as general a form as possible without resorting to unnecessary numerical analysis in specific designs.
We assume that the pump propagates only in the inner-cladding (and the core) and has a uniform intensity of $I_p$, which 
can also be a function of $z$. $P_p=\pi b^2 I_p$ is the total pump power.
This assumption is usually valid if the pump laser is sufficiently scrambled to maintain its uniformity in the transverse plane.
We also define the pump overlap factor with the core as $\Gamma:=a^2/b^2$. In some DCF designs, the circular symmetry of the inner-cladding is broken to help maintain 
the transverse uniformity of the pump intensity along the fiber. These technicalities do not affects our general conclusions.
Both the signal intensity (nearly Gaussian) and power are assumed to generally depend on the longitudinal coordinate, $z$, along the fiber. 
We assume that the fiber temperature is constant in time; therefore, the steady state heat equation can be used to determine the temperature distribution, $T(x,y,z)$:  
\begin{align}
\label{Eq:heatdiff}
\nabla\cdot(\kappa\nabla T)+q=0,
\end{align}
where the heat source density $q(x,y,z)$ is the thermal energy deposited per second at the location $(x,y,z)$ inside the fiber and is in units of ${\rm W/m}^3$
and $\kappa(x,y,z)$ is the local thermal conductivity~\cite{book}. 
Equation~\ref{Eq:heatdiff}, in general, can only be solved numerically. Here, we make another simplifying assumption that the temperature gradient in the $z$-direction 
varies very slowly with $z$ ($\partial^2_zT\approx 0$). 
We will discuss the validity of this assumption later in Appendix A. Therefore, considering the cylindrical symmetry of the problem and ignoring the $\partial^2_z T$ term 
in Eq.~\ref{Eq:heatdiff}, we arrive at the following differential equation:
\begin{align}
\label{Eq:heatdiff2}
\kappa\dfrac{\partial^{2}T}{\partial\rho^{2}}+\left(\dfrac{\kappa}{\rho}+\dfrac{\partial\kappa}{\partial\rho}\right)\dfrac{\partial T}{\partial\rho}+q=0,
\end{align}
where $\rho$ is the radial coordinate. Both $T$ and $q$ are in general functions of $z$ as well as $\rho$; however, in the following discussion, 
we drop their explicit $z$-dependence when writing the equations for simplicity, but their $z$-dependence is always implicitly assumed (see Appendix A).   
Because of the geometry of the fiber in Fig.~\ref{fig:fiber-temp}, $\kappa(\rho)$ is piece-wise constant; therefore,
the term proportional $\partial_\rho\kappa$ vanishes in each segment. Moreover, we assume that $q$ is piece-wise constant (this assumption is justified in Appendix B); 
therefore, the general solution to the second-order ordinary differential equation~\ref{Eq:heatdiff2} in each radial segment of the fiber is given by:
\begin{align}
\label{Eq:solgeneral}
T(\rho)={\mathcal C}_1+{\mathcal C}_2\log(\rho^2)-\dfrac{q}{4\kappa}\rho^2,
\end{align}
where ${\mathcal C}_1$ and ${\mathcal C}_2$ are constants of integration. 

In subsection~\ref{sec:formulation}\ref{sec:define}, we summarize the definitions of the main parameters used in this paper for the temperature profiles, and 
in subsection~\ref{sec:formulation}\ref{sec:solution}, we present the solutions that provide the temperature distributions inside the optical fiber.
In subsection~\ref{sec:formulation}\ref{sec:thermal}, we will present and justify the values of the thermal parameters used in the paper, and in 
subsection~\ref{sec:formulation}\ref{sec:scalars}, we will elaborate on the order-one scalar coefficients that appear in equations similar to Eq.~\ref{Eq:1stresult}.
\subsection{Definition of the parameters}
\label{sec:define}
The following parameters will be used in the solutions of Eq.~\ref{Eq:heatdiff2}:
\begin{packedit1}
\item $q_a$ is the uniform heat density inside the core ($0\le\rho\le a$) due to the quantum defect, as well as the parasitic absorption of the signal and the pump.
\item $q_b$ is the uniform heat density inside the inner-cladding ($a<\rho\le b$) due to the parasitic absorption of the pump.
\item No heat is generated in the outer-cladding region ($b<\rho\le c$) in which no signal or pump propagates.
\item $\kappa_a$, $\kappa_b$, and $\kappa_c$ are the thermal conductivities in the core, inner-cladding, and outer-cladding regions. Their values are 
assumed to be uniform in each region. 
\item The thermal characteristic length scales in each radial fiber segment is defined as $\mathcal{D}_i=4\kappa_i/H$, where $i=a,b,c$.
\item The total linear heat density generated in the core is given by $Q_a=q_a\pi a^2$. 
Similarity, the total linear heat density generated in the 
inner-cladding is given by $Q_b=q_b\pi (b^2- a^2)$. We also define $Q_p=q_b\pi b^2$, which represents the total generated linear heat density due to the parasitic absorption of the pump.
\item We define the following temperatures based on $T(\rho)$: $T_0=T(0)$, $T_a=T(a)$, $T_b=T(b)$, $T_c=T(c)$, and $T_\infty$ as the ambient outside temperature.
We also define the following temperature variation parameters: $\delta T_a=T_0-T_a$, $\delta T_b=T_a-T_b$, $\delta T_c=T_b-T_c$, and $\Delta T=T_c-T_\infty$.
\item We use the following geometrical parameters: $\Gamma=a^2/b^2$, $\eta=\ln (c^2/b^2)$, and $D=2c$. 
\end{packedit1} 
\subsection{Solution of the temperature equation}
\label{sec:solution}
In the core, the temperature must be finite everywhere including at $\rho=0$; therefore,
${\mathcal C}_2$ from Eq.~\ref{Eq:solgeneral} must vanish in the core. Moreover, the temperature and the radial heat flux must be continuous at each layer. 
We remind that the radial heat flux is given by $-\kappa\partial_\rho T(\rho)$, where $\kappa$ is the relevant thermal conductivity in each region.
The result is the following temperature profile inside the fiber at each layer:
\begin{align}
\label{Eq:allsolutions}
&T_0-T(\rho)=\dfrac{q_a\,\rho^2}{4\kappa_a},\ \ 0\le\rho\le a,\\
\nonumber
&T_a-T(\rho)=\dfrac{(q_a-q_b)a^2}{4\kappa_b}\ln(\dfrac{\rho^2}{a^2})+\dfrac{q_b(\rho^2-a^2)}{4\kappa_b},\ \ a<\rho\le b,\\
\nonumber
&T_b-T(\rho)=\dfrac{q_a\, a^2+q_b(b^2-a^2)}{4\kappa_c}\ln(\dfrac{\rho^2}{b^2}),\ \ b<\rho\le c.
\end{align}
The temperature variations can be obtained as
\begin{subequations}
\begin{align}
\label{Eq:dTadef}
&(4\pi\kappa_a)\delta T_a=Q_a,\\
\label{Eq:dTbdef}
&(4\pi\kappa_b)\delta T_b=(Q_a+Q_b-Q_p)(-\ln\Gamma)+Q_b,\\
\label{Eq:dTcdef}
&(4\pi\kappa_c)\delta T_c=(Q_a+Q_b)\eta.
\end{align}
\end{subequations}

We assume convective boundary condition for the outer surface of the fiber: 
\begin{align}
-\kappa_c\dfrac{\partial T(\rho)}{\partial\rho}\Big|_c=H(T_c-T_\infty).
\end{align}
Therefore, the temperature difference between the fiber surface $T_c$ and ambient outside temperature $T_\infty$ is given by
\begin{align}
\label{Eq:DtQaQb}
\Delta T=T_c-T_{\infty}=\dfrac{Q_a+Q_b}{2\pi c\,H}.
\end{align}

We are now ready to simplify the previous equations and the following two relationship will be useful in the process:
\begin{align}
\label{Eq:QaQbdefine}
Q_b=(1 - \Gamma)Q_p,\qquad Q_a=(\Gamma + \gamma) Q_p.
\end{align}
The meaning of the coefficient $\gamma$ will become clear shortly (in subsection~\ref{sec:formulation}\ref{sec:scalars}), but for now, Eq.~\ref{Eq:QaQbdefine}
can be used as the definition of $\gamma$. After a few lines of algebra, we arrive at: 
\begin{subequations}
\begin{align}
\label{Eq:finaldTa}
\dfrac{\delta T_a}{\Delta T}&=\Big[\dfrac{\Gamma+\gamma}{1+\gamma}\Big]\,\dfrac{D}{\mathcal{D}_a},\\
\label{Eq:finaldTb}
\dfrac{\delta T_b}{\Delta T}&=\Big[\dfrac{1-\Gamma}{1+\gamma}+\dfrac{\gamma}{1+\gamma}(-\ln\Gamma)\Big]\,\dfrac{D}{\mathcal{D}_b},\\
\label{Eq:finaldTc}
\dfrac{\delta T_c}{\Delta T}&=\eta\,\dfrac{D}{\mathcal{D}_c}.
\end{align}
\end{subequations}
Equations~\ref{Eq:finaldTa},~\ref{Eq:finaldTb}, and ~\ref{Eq:finaldTc} are the main results of this paper and will be analyzed 
in detail in the following discussions. Note that the total temperature change from the center of the core to the surface of the 
fiber is given by $\delta T_a+\delta T_b+\delta T_c$.
\subsection{Relevant thermal parameters}
\label{sec:thermal}
In this subsection, we present the values of the relevant thermal parameters that will be used in the rest of this paper. 
The core and the inner-cladding of a typical DCF in fiber laser applications are made from fused silica. For both of these
regions, we assume a uniform thermal conductivity across the fiber because the light doping of various elements in the core and inner-cladding 
do not change the value of $\kappa$ in the host glass, substantially. For fused silica at room temperature, we have 
$\kappa=1.38\,{\rm W/(m.K)}$~\cite{Heraeus}, which is the value we will use in the subsequent analysis. However, the value of 
$\kappa$ increases with temperature and can reach $\approx 1.65\,{\rm W/(m.K)}$ at $320^{\circ}$C~\cite{Stabler}.
For the outer-cladding, low-index Acrylate and Polyimide polymers are commonly used among many other polymers. While 
Acrylate can reliably withstand temperatures as high as $120^{\circ}$C, Polyimide coating is preferred for high
power applications because it remains reliable to as high as $300^{\circ}$C in continuous operation~\cite{Huang}. 
For Polyimide, we use $\kappa\approx 0.276\,{\rm W/(m.K)}$~\cite{matweb}.

The value of the convective heat transfer coefficient $H$ depends on the choice of fluid and its speed. We
use $H\approx 92\,{\rm W/(m^{2}.K)}$ for a high-speed-air-fan-cooled fiber and $H\approx 920\,{\rm W/(m^{2}.K)}$
for a moderate-flow-speed-water-cooled fiber~\cite{HeatTransfer}. 

\begin{table}[htp]
   \caption{The thermal characteristic length scale defined as $\mathcal{D}_i=4\kappa_i/H$ for the choices of material and convection fluid. The 
            subscript $sa$ stands for ``silica'' material and ``air'' cooling and so on.}
\begin{center}
\label{kappaH}
 \renewcommand{\arraystretch}{1.3}
\begin{tabular}{ | p{29mm} | p{27mm} | p{28mm} |}
      \hline
      {\small }    & {\small silica:\qquad\qquad\qquad\qquad $\kappa=1.38\,{\rm W/(m.K)}$}  & {\small Polyimide:\qquad\qquad\quad $\kappa=0.276\,{\rm W/(m.K)}$}   \\
      \hline
      {\small forced-air-cooling $H\approx 92\,{\rm W/(m^{2}.K)}$}       &    $\mathcal{D}_{sa}=6\,{\rm cm}$  &   $\mathcal{D}_{pa}=1.2\,{\rm cm}$                 \\
      \hline
      {\small forced-water-cooling $H\approx 920\,{\rm W/(m^{2}.K)}$}    &    $\mathcal{D}_{sw}=6\,{\rm mm}$  &   $\mathcal{D}_{pw}=1.2\,{\rm mm}$                 \\
      \hline
\end{tabular}
\end{center}
\end{table}
It is clear that $\mathcal{D}_a$ and $\mathcal{D}_b$ can take the value of $\mathcal{D}_{sa}$ or $\mathcal{D}_{sw}$ depending on the choice of the cooling fluid,
while $\mathcal{D}_b$ can take the value of $\mathcal{D}_{pa}$ or $\mathcal{D}_{pw}$. These thermal characteristic length scales must be compared with the outer 
diameter of the fiber as they appear in the form of a ratio in Eqs.~\ref{Eq:finaldTa},~\ref{Eq:finaldTb}, and ~\ref{Eq:finaldTc}. The outer 
diameter of the fiber $D$ (including the polymer coating) typically ranges from $\approx 250\,\mu{\rm m}$ to $\approx 500\,\mu{\rm m}$, so $D/\mathcal{D}_{pw}$
ranges from 2.5 to 5, $D/\mathcal{D}_{pa}$ ranges from 25 to 50,  $D/\mathcal{D}_{sw}$ ranges from 12 to 25, $D/\mathcal{D}_{sa}$ ranges from 120 to 500. 
These ratios affect the values of temperature variations in Eqs.~\ref{Eq:finaldTa},~\ref{Eq:finaldTb}, and ~\ref{Eq:finaldTc}.
\subsection{Scalar coefficients}
\label{sec:scalars}
In this subsection, we examine the scalar coefficients that appear in Eqs.~\ref{Eq:finaldTa},~\ref{Eq:finaldTb}, and ~\ref{Eq:finaldTc}, behind the 
ratio of the length scales $D/\mathcal{D}_i$:
\begin{subequations}
\begin{align}
\label{Eq:finalXa}
\mathfrak{X}_a&=\dfrac{\Gamma+\gamma}{1+\gamma},\\
\label{Eq:finalXb}
\mathfrak{X}_b&=\dfrac{1-\Gamma}{1+\gamma}+\dfrac{\gamma}{1+\gamma}(-\ln\Gamma),\\
\label{Eq:finalXc}
\mathfrak{X}_c&=\eta.
\end{align}
\end{subequations}
We will argue that these coefficients are all order one scalars, so in each case $\delta T/\Delta T$ is primarily set by the ratio $D/\mathcal{D}_i$.

In a conventional cladding-pumped fiber laser or amplifier, the primary sources of heating are from the quantum defect in the core of the fiber and the parasitic absorption of 
both the signal and the pump. Based on our assumptions, the heating due to the quantum defect happens uniformly in the core with the linear heat density of $Q_{qd}$, which only contributes to 
$Q_a$. The linear heat density due to the parasitic absorption of the signal is given by $Q_{as}=\alpha_s P_s$ ($P_s$ is the total signal power),  which only contributes to $Q_a$, as well.
The linear heat density due to the parasitic absorption of the pump is given by $Q_{ap}=\alpha_p P_p$, a fraction of which, $\Gamma Q_{ap}$, contributes to $Q_a$ and 
the rest, $(1-\Gamma)Q_{ap}$, is deposited in the inner-cladding and contributes to $Q_b$. Here, $\alpha_s$ and and $\alpha_p$ are 
the parasitic absorption coefficients of the signal and the pump, respectively. Using these definitions, we obtain:
\begin{align}
\label{Eq:DCFQaQb}
Q_a=Q_{qd}+Q_{as}+\Gamma Q_{ap},\qquad
Q_b=(1-\Gamma)Q_{ap}. 
\end{align}
Using Eq.~\ref{Eq:QaQbdefine} and the definitions presented in Eq.~\ref{Eq:DCFQaQb}, it can be shown that 
\begin{align}
\label{Eq:gamma}
\gamma=\dfrac{Q_{qd}+Q_{as}}{Q_{ap}}. 
\end{align}
Equation~\ref{Eq:gamma} makes the meaning of the parameter $\gamma$ more clear: $\gamma$ is the ratio of the sum of the quantum defect linear heat density and the signal parasitic 
absorption, both of which are deposited in the core, to the total parasitic heat generation in the fiber due to the pump. Note that
$\gamma$ appears in Eqs.~\ref{Eq:finalXa},~\ref{Eq:finalXb}, and~\ref{Eq:finalXc} in the form of $1/(1+\gamma)$ and $\gamma/(1+\gamma)$, both of which are always between 0 and 1, i.e., 
$0\le 1/(1+\gamma) \le 1$ and $0\le \gamma/(1+\gamma) \le 1$ for $0\le\gamma<\infty$; therefore, their finite values (bounded from above) set their contribution levels 
to $\mathfrak{X}_a$ and $\mathfrak{X}_b$.

From these arguments, we find that depending on the relative size of the contributions from $Q_{qd}$, $Q_{as}$, and $Q_{ap}$, which set the value of $\gamma$, we have
the following acceptable ranges for the scalar coefficients:
\begin{align}
\label{Eq:ranges}
\Gamma<\mathfrak{X}_a<1,\qquad 1-\Gamma<\mathfrak{X}_b<(-\ln\Gamma),\qquad \mathfrak{X}_c=\eta.
\end{align}
The upper limit in each case is obtained for $\gamma \gg 1$, which is usually the case for conventional fiber lasers and amplifiers, where $Q_{qd}$ is much higher than
$Q_{as}$ and $Q_{ap}$ (see Eq.~\ref{Eq:gamma}). However, in some modern high-power fiber amplifiers where the quantum defect is lowered~\cite{Li}, or when the amplifier operates 
in a nearly radiation-balanced regime or for radiation-balanced lasers~\cite{bowman2010minimizing,bowman2016low,mobini2018thermal,mobiniCoreCladding}, $\gamma<1$ 
and a value closer to the lower limit in Eq.~\ref{Eq:ranges} may be applicable. Note that $\mathfrak{X}_c=\eta$ does not deviate much from unity in conventional DCFs. 
\section{Examples}
\label{sec:examples}
In the following, we will explore three examples of Yb-doped optical fibers from Thorlabs Incorporated, where the relevant fiber parameters are given
in Table~\ref{silicazblan}. We note that in all three cases, Thorlabs Incorporated reports Acrylate polymer coating, which can only withstand temperatures up to 
$120^{\circ}$C. For our analysis, we will assume Polyimide coating because its temperature can go as high as $300^{\circ}$C; this choice is justified because our discussions
are primarily aimed at high power laser operations. For the rest of the discussion, we assume that the fiber surface is heated to $320^{\circ}$C, so 
$\Delta T=300^{\circ}$C, where $T_\infty=20^{\circ}$C is assumed. For Acrylate polymer coating where $\Delta T\approx 100^{\circ}$C, all temperature values obtained
below must be divided by a factor of three. We emphasize that {\em Fiber1} is practically a single-cladding fiber and is not commonly used in high-power operation. 
However, we have included this fiber here to show that the analysis in this paper can also apply to this special case, noting that the results for this fiber
are somewhat less interesting than those for {\em Fiber2} and {\em Fiber3}, which are high-power DCFs.
\begin{table}[htp]
   \caption{The relevant fiber parameters for Eqs.~\ref{Eq:finaldTa},~\ref{Eq:finaldTb}, and~\ref{Eq:finaldTc}.}
\begin{center}
\label{silicazblan}
 \renewcommand{\arraystretch}{1.3}
\begin{tabular}{ | p{10mm} | p{30mm} | p{12mm} | p{12mm} | p{12mm} |}
      \hline
      {\small name}          & {\small fiber ID}                 & $2a (\mu{\rm m})$    & $2b (\mu{\rm m})$    & $2c (\mu{\rm m})$   \\
      \hline
      {\small\em Fiber1}    & {\small YB1200-4/125}    & 4                    & 125                    & 245                       \\
      \hline
      {\small\em Fiber2}    & {\small YB1200-10/125DC} & 10                   & 125                    & 245                       \\
      \hline
      {\small\em Fiber3}    & {\small YB1200-20/400DC} & 20                   & 400                    & 520                       \\ \hline
\end{tabular}
\end{center}
\end{table}

In Tables~\ref{tableGammaeta}, and~\ref{tableDvD}, we calculate the values of the geometrical parameters and ratios of the thermal length scale to the outer diameter of
for fiber for Eqs.~\ref{Eq:finaldTa},~\ref{Eq:finaldTb}, and~\ref{Eq:finaldTc}. In Table~\ref{tabledeltat}, we use the information in Eqs.~\ref{Eq:finalXa},~\ref{Eq:finalXb}, and~\ref{Eq:finalXc}
to estimate the temperature variations $\delta T_a$, $\delta T_b$, and $\delta T_c$, in the core, inner-cladding, and outer-cladding of each fiber, respectively. The results are reported
for both air-cooling and water-cooling, and for a range that depends on the value of $\gamma$ as discussed earlier. In Table~\ref{tablelinearheat},
we use Eq.~\ref{Eq:DtQaQb} to calculate the total linear heat density $Q_a+Q_b$ that must be deposited inside the optical fiber to heat the surface temperature by $\Delta T=300^{\circ}$C
relative to the ambient; of course, a much larger heat deposit is needed to reach the same level of $\Delta T$ for water cooling relative to air cooling.  

We next consider the temperature changes for a nominal value of $Q_a+Q_b=50\,{\rm W/m}$, which is a typical value used in modern high-power fiber amplifiers~\cite{Beier}. 
Note that in the previous analysis, we fixed $\Delta T=300^{\circ}$C, but here we allow it to vary and instead fix $Q_a+Q_b$.
The corresponding temperature ranges are reported in Table~\ref{table50Wm} and they are the same for air-cooling and water-cooling, because $H$ cancels out if 
$\Delta T$ from Eq.~\ref{Eq:DtQaQb} is used in Eqs.~\ref{Eq:finaldTa},~\ref{Eq:finaldTb}, and~\ref{Eq:finaldTc}. The total temperature change is also reported as 
$\sum\delta T_i$, which is the quantity measured in Ref.~\cite{Beier}. As will be noted in section~\ref{sec:examples}\ref{sec:gamma}, the upper limit values coming from
$\gamma\gg 1$ correspond to most conventional systems for which the heat density due to quantum defect overwhelms other sources of heat. In the limit of
$\gamma\gg 1$, we obtain
\begin{align}
\label{Eq:maxtemp}
\sum\delta T_i=\left[\dfrac{1}{4\pi\kappa_a}+\dfrac{(-\ln\Gamma)}{4\pi\kappa_b}+\dfrac{\eta}{4\pi\kappa_c}\right](Q_a+Q_b).
\end{align}
To make a comparison with the results reported in Ref.~\cite{Beier} where $2a=25\,\mu{\rm m}$ and $2b=400\,\mu{\rm m}$, if we consider the case 
of $Q_a+Q_b=35\,{\rm W/m}$, we obtain $\sum\delta T_i=22^{\circ}$C, which is in close agreement with their direct temperature measurement. 
Of course, if the measured core temperature is only the average value (as is the case in Ref.~\cite{Beier}), $1/4\pi\kappa_a$ in Eq.~\ref{Eq:maxtemp} 
must be replaced with $1/8\pi\kappa_a$ because $\int_0^a \rho^2 \rho d\rho/(a^2 \int_0^a \rho d\rho)=1/2$ (see Eq.~\ref{Eq:allsolutions}), which
results in $\sum\delta T_i=21^{\circ}$C.

\begin{table}[h!]
   \caption{The values of the geometrical parameters.}
\begin{center}
\label{tableGammaeta}
 \renewcommand{\arraystretch}{1.3}
\begin{tabular}{ | p{10mm} | p{12mm} | p{12mm} | p{12mm} |}
      \hline
                             & $\Gamma$ & $-\ln\Gamma$ & $\eta$  \\
      \hline
      {\small\em Fiber1}     &  0.0010   & 6.88           & 1.35  \\
      \hline
      {\small\em Fiber2}     &  0.0064   & 5.05           & 1.35  \\
      \hline
      {\small\em Fiber3}    &  0.0025    & 5.99           & 0.525 \\ \hline
\end{tabular}
\end{center}
\end{table}
\begin{table}[h!]
   \caption{The ratio of the thermal length scale to the outer diameter.}
\begin{center}
\label{tableDvD}
 \renewcommand{\arraystretch}{1.3}
\begin{tabular}{ | p{10mm} | p{13mm} | p{13mm} | p{13mm} | p{13mm} |}
      \hline
                             & $\mathcal{D}_{sa}/D$ & $\mathcal{D}_{sw}/D$ & $\mathcal{D}_{pa}/D$ & $\mathcal{D}_{pw}/D$  \\
      \hline
      {\small\em Fiber1}     &  123 & 12.2 & 24.5 & 2.45  \\
      \hline
      {\small\em Fiber2}     &  123 & 12.2 & 24.5 & 2.45 \\
      \hline
      {\small\em Fiber3}     & 57.7 & 5.77 & 11.5 & 1.15 \\ \hline
\end{tabular}
\end{center}
\end{table}
\begin{table}[h!]
   \caption{Temperature variation parameter ranges in $^{\circ}$C  if $\Delta T=300^{\circ}$C.}
\begin{center}
\label{tabledeltat}
 \renewcommand{\arraystretch}{1.3}
\begin{tabular}{ | p{10mm} | p{13mm} | p{10mm} | p{6mm} || p{13mm} | p{10mm} | p{6mm} |}
      \hline
                             & \multicolumn{3}{ c|| }{air-cooling} & \multicolumn{3}{ c | }{water-cooling} \\
      \hline
                             & $\delta T_a$ & $\delta T_b$ & $\delta T_c$ & $\delta T_a$ & $\delta T_b$ & $\delta T_c$  \\
      \hline
      {\small\em Fiber1}     &  {\small 0.003-2.5} & \small{2.5-17} & {\small 17} & {\small 0.025-25} & {\small 25-169} & {\small 165}  \\
      \hline
      {\small\em Fiber2}     &  {\small 0.016-2.5} & \small{2.4-12} & {\small 17} & {\small 0.16-25} & {\small 24-124} & {\small 165}  \\
      \hline
      {\small\em Fiber3}     & {\small 0.013-5.2} & \small{5.2-31} & {\small 14} & {\small 0.13-52} & {\small 52-312} & {\small 136}   \\ \hline
\end{tabular}
\end{center}
\end{table}
\begin{table}[h!]
   \caption{The total linear heat density $Q_a+Q_b$ needed to raise the surface temperature of the fiber relative to the ambient by $\Delta T=300^{\circ}$C.}
\begin{center}
\label{tablelinearheat}
 \renewcommand{\arraystretch}{1.3}
\begin{tabular}{ | p{10mm} | c | c |}
      \hline
                             & air-cooling (W/m) & water-cooling (W/m)   \\
      \hline
      {\small\em Fiber1}     &  43   & 425         \\
      \hline
      {\small\em Fiber2}     &  43   & 425          \\
      \hline
      {\small\em Fiber3}     &  90   & 902          \\ \hline
\end{tabular}
\end{center}
\end{table}
\begin{table}[h!]
   \caption{Temperature variation parameter ranges in $^{\circ}$C  if total linear heat density is $Q_a+Q_b=50\,{\rm W/m}$.  
            $\delta T_a$, $\delta T_b$, and $\delta T_c$ are the same for both air-cooling and water cooling. We also define
            $\sum\delta T_i=\delta T_a+\delta T_b+\delta T_c$, which is the total temperature change inside the fiber.
            $\Delta T_{ac}$ is the surface temperature relative to the ambient for air-cooling and $\Delta T_{aw}$ is for water cooling.}
\begin{center}
\label{table50Wm}
 \renewcommand{\arraystretch}{1.3}
\begin{tabular}{ | p{10mm} | p{13mm} | p{10mm} | p{6mm} | p{9mm} | p{9mm} | p{9mm} |}

      \hline
                             & $\delta T_a$ & $\delta T_b$ & $\delta T_c$ & $\sum\delta T_i$ & $\Delta T_{ac}$ & $\Delta T_{wc}$  \\
      \hline
      {\small\em Fiber1}     &  {\small 0.003-2.9} & \small{2.9-20} & {\small 19} & {\small 22-42} & {\small 353} & {\small 35.3}  \\
      \hline
      {\small\em Fiber2}     &  {\small 0.019-2.9} & \small{2.9-15} & {\small 19} & {\small 22-37} & {\small 353} & {\small 35.3}  \\
      \hline
      {\small\em Fiber3}     & {\small 0.007-2.9} & \small{2.9-17} & {\small 8} & {\small 11-28} & {\small 166} & {\small 16.6}   \\ \hline
\end{tabular}
\end{center}
\end{table}
\subsection{Can the temperature rise result in multimode operation?}
The results presented so far focus mainly on the transverse temperature variations in fused silica optical fibers. In practice, the principal optical quantity of interest is the induced change in the refractive index due to the temperature change. The change in the refractive index is related to the change in the temperature by the thermo-optic coefficient, $dn/dT$. For fused silica, the thermo-optic coefficient is reported at 546\,nm to be $11.3\times 10^{-6}\,{\rm K}^{-1}$~\cite{Rocha}, and does not vary substantially with the wavelength. The V-number of a step-index optical fiber for a small core-cladding index contrast of $\Delta$ is $V\approx 2\pi a \sqrt{2n\Delta}/\lambda$, where $a$ is the core radius, $n\approx 1.5$ is the average refractive index, and $\lambda$ is the optical wavelength.  The single-mode cut-off is at $V\approx 2.405$. If a temperature rise results in a substantial change in the value of $\Delta$, it can turn a single-mode fiber to multimode. Using the definition of the V-number, we can show that $\delta V/V=\delta \Delta/(2\Delta)$, where $\delta \Delta$ is the change in the core-cladding index difference. For a single-mode large-core fiber of $V\approx 2.3$ with a $30\,\mu{\rm m}$ core diameter at $\lambda\approx 1\,\mu{\rm m}$, $\Delta\approx 2\times 10^{-4}$. Assuming a maximum tolerable change in the V-number of 20\%, the maximum acceptable $\delta\Delta$ is $8\times 10^{-5}$. We can approximate $\delta\Delta\approx (dn/dT)\delta T_a$, given that $\delta T_a$ sets the scale for the core-cladding temperature variation, resulting in a maximum acceptable value of $\delta T_a\approx 8^{\circ}$C. Of course, the temperature profile in the core is not of a top-hat form and decreases quadratically, so this analysis slightly underestimates the $\delta T_a$ required for the fiber core to support multiple modes.

The results presented for the value of $\delta T_a$ in Table~\ref{tabledeltat} indicate that for air-cooling and $\Delta T= 300^{\circ}$C, 
the temperature variation range in the core is quite small and the fiber is unlikely to transition from single-mode to multimode. 
However, in a water-cooled system, the value of $\delta T_a$ can be as high as $25^{\circ}$C for {\em Fiber1} and {\em Fiber2} and as 
high as $50^{\circ}$C for {\em Fiber3}, which can clearly result in a multimode core in the high-heat and $\gamma\gg 1$ (conventional) operation. 
\subsection{Estimation of the $\gamma$ parameter}
\label{sec:gamma}
We already noted in Eq.~\ref{Eq:gamma} that $\gamma$ is the ratio of the sum of the quantum defect linear heat density and the signal parasitic 
absorption to the total parasitic heat generation in the fiber due to the pump. We can obtain an estimate of this parameter if we assume for example
an amplifier set-up in the forward pumping configuration, where at the input the signal power can be neglected compared with the pump power.
In this case, $Q_{qd}\approx \alpha_r P_p \delta\lambda/\lambda_s$ (see Appendix B for the derivation). 
$\alpha_r$ is the resonant pump absorption coefficient and is given by $N_t\sigma^a_p\Gamma$, where 
$N_t$ is total Yb ion dopant density and $\sigma^a_p$ is the absorption cross section of the pump. $\delta\lambda=\lambda_s-\lambda_p$ is the difference between 
the wavelengths of the signal and the pump. Noting that $Q_{ap}=\alpha_p P_p$ and $Q_{as}\approx 0$, so we have
\begin{align}
\label{Eq:gamma2}
\gamma\approx \dfrac{\alpha_r}{\alpha_p}\times\dfrac{\delta\lambda}{\lambda_s}. 
\end{align}

This is an interesting result and gives an estimate on the value of $\gamma$, independent of the pump power. For our analysis, we consider a nominal value of 
$\alpha_p\approx 15\,{\rm dB/km}$, which is reasonable for conventional high-power fiber lasers. For pumping at the peak absorption wavelength $\lambda_p=976\,{\rm nm}$, 
the nominal values of $\alpha_r$ for {\em Fiber1}, {\em Fiber2}, and {\em Fiber3} are $1200\,{\rm dB/m}$, $7.4\,{\rm dB/m}$, and $3\,{\rm dB/m}$, respectively.
For $\lambda_s\approx 1064\,{\rm nm}$, $\delta\lambda/\lambda_s\approx 0.1$, so the corresponding values of $\gamma$ are approximately $8000$, $50$, and $20$, respectively.
If the same fibers are pumped at $\lambda_p=920\,{\rm nm}$, the measured values of $\alpha_r$ for {\em Fiber1}, {\em Fiber2}, and {\em Fiber3} 
would be $280\,{\rm dB/m}$, $1.7\,{\rm dB/m}$, and $0.7\,{\rm dB/m}$, respectively. Given that $\delta\lambda/\lambda_s\approx 0.14$, 
the corresponding values of $\gamma$ are approximately $2520$, $15.3$, and $6.3$, respectively.

However, for a low quantum defect fiber amplifier~\cite{Li} pumped at $\lambda_p=1018\,{\rm nm}$, the nominal value of $\alpha_r$ would be 10 times smaller than that for
$\lambda_p=920\,{\rm nm}$ due to a smaller absorption cross section of the pump. In this case, $\delta\lambda/\lambda_s\approx 0.05$, and
the corresponding values of $\gamma$ are approximately $93$, $0.57$, and $0.23$, respectively. Note the small value of $\gamma$ for {\em Fiber3}, which is often used
in high-power operation (unlike {\em Fiber1}, which is a single-cladding fiber). The value of $\gamma$ can become even lower in the 
nearly radiation-balanced regime~\cite{bowman2010minimizing,bowman2016low,mobini2018thermal,mobiniCoreCladding}.
\section{Summary and Conclusion}
\label{sec:summary}
The analytical expressions for temperature variations are simple and can be used for quick estimation of the temperature distributions inside the optical fiber. 
The inclusion of the heat generation in the cladding is essential for modern high-power fiber lasers and amplifiers, 
where the quantum defect is lowered, or when the amplifier operates in a nearly radiation-balanced regime, or for radiation-balanced lasers.
In all these cases, the heat generated due to the parasitic absorption of the high-power pump in the inner-cladding can be considerable and must be included in the analysis. 
The analytical expressions can be used for a wide range of conventional DCF-based systems. A single parameter 
$0\ll\gamma<\infty$ allows one to interpolate between the case where the quantum defect heating is dominant to when the parasitic absorption heating is 
comparable in size or is the dominant source of heating.

For the numerical analysis of the analytical expressions, we consider the maximum tolerable surface temperature of the fiber relative to the ambient to be 
$\Delta T=300^{\circ}$C to protect the Polyimide coating that is the polymer of choice in high-temperature operation. Our results show that for air-cooled DCFs
with $\Delta T\lesssim 300^{\circ}$C, the core temperature rises only in the range of $0-5^{\circ}$C relative to the inner-cladding; however, for 
water-cooled DCFs, the core temperature can be higher than the inner-cladding by as much as $50^{\circ}$C, potentially resulting in a change from the 
single-mode core to the multimode cre due to the thermo-optic effect.

Last but not least, in Appendix C, we summarize the equations that are most useful for direct comparison with experiments.
\section*{Appendix A: Longitudinal variation of the temperature}
In this Appendix, we would like to justify the absence of the $\partial^2_z T$ term in our analysis based on Eq.~\ref{Eq:heatdiff2}.
Let's consider a situation where $q$ is Eq.~\ref{Eq:heatdiff} can be expressed as $q(\rho,z)=\tilde{q}(\rho)\,f(z)$. This separable form,
while very convenient in our analysis, can be fully justified if only one of the heating sources, $Q_{qd}$, $Q_{as}$, or $Q_{ap}$,
is the dominant one. However, our discussion captures the essence of why the $\partial^2_z T$ term can be ignored, regardless. 

The analysis presented in this paper means that the temperature profile has a $\rho$-dependence 
subject to the form of Eq.~\ref{Eq:heatdiff2} with $\tilde{q}(\rho)$ as the heat source, and a $z$-dependence of the form $f(z)$. 
In other words, $T(\rho,z)\approx \widetilde{T}(\rho)f(z)$, where
\begin{align}
\label{Eq:heatdiff3}
\frac{\partial^{2}\widetilde{T}(\rho)}{\partial\rho^{2}}+\frac{1}{\rho}\frac{\partial \widetilde{T}(\rho)}{\partial\rho}+\dfrac{\tilde{q}(\rho)}{\kappa}=0.
\end{align}
Without making any approximations, the full form of the temperature profile can be expressed as 
\begin{align}
\label{Eq:Tfull}
T(\rho,z)=\widetilde{T}(\rho)f(z)+\tau(\rho,z),
\end{align}
where $\tau(\rho,z)$ should be negligible if our approximations hold. In other words, the size of $\tau(\rho,z)$ characterizes the relative 
importance of keeping the $\partial^2_z T$ term in Eq.~\ref{Eq:heatdiff}.

If we substitute $T(\rho,z)$ from Eq.~\ref{Eq:Tfull} in Eq.~\ref{Eq:heatdiff}, while considering Eq.~\ref{Eq:heatdiff3}, we arrive at
\begin{align}
\label{Eq:heatdiff4}
\nabla^2\,\tau(\rho,z)+\widetilde{T}(\rho)\,\partial_z^2f(z)=0.
\end{align}
In Eq.~\ref{Eq:heatdiff4}, the term $\partial^2_z f(z)$ can be approximated into the form of $f(z)/\widetilde{L}$, where $\widetilde{L}$ is a length-scale
on the order of the full length of the optical fiber. This can be understood for example if $f(z)\sim \exp(-\widetilde{\alpha}\,z)$,
where $\widetilde{\alpha}$ can be, e.g., the absorption coefficient of the pump, and the pump power is almost entirely absorbed over the full
length of the fiber laser. If the pump power is not fully absorbed, then $\widetilde{L}$ can be even larger than the length of the optical fiber.
Next, looking at Eq.~\ref{Eq:allsolutions} reveals that the radial temperature profile of the fiber has, generally speaking, the form of
$\widetilde{T}(\rho)\sim \tilde{\rho}^2\,\tilde{q}(\rho)/4\kappa$, where $\tilde{\rho}^2$ is a length scale comparable to the radius of the fiber.
Therefore, we can approximate Eq.~\ref{Eq:heatdiff4} as
\begin{align}
\label{Eq:heatdiff5}
\nabla^2\tau(\rho,z)+\dfrac{\tilde{\rho}^2}{\widetilde{L}^2}\dfrac{\tilde{q}(\rho)f(z)}{\kappa}=0.
\end{align}
Comparing Eq.~\ref{Eq:heatdiff5} with Eq.~\ref{Eq:heatdiff3}, it can be readily observed that 
\begin{align}
\tau(\rho,z)\sim \dfrac{\tilde{\rho}^2}{\widetilde{L}^2}\widetilde{T}(\rho)f(z),
\end{align}
therefore, $\tau(\rho,z)$ is smaller than $\widetilde{T}(\rho)f(z)$ by the factor of $\tilde{\rho}^2/\widetilde{L}^2$, which is usually 4 orders of magnitude 
or more.

In summary, $\partial^2_z T$ term can be ignored unless longitudinal variations in heat deposit in the fiber occur at scales comparable to the fiber diameter, 
which is hardly conceivable in steady-state.
\section*{Appendix B: Validity of Top hat Assumption}
In the main text of the paper, we assume that the heat density $q$ is piece-wise constant, which serves as a convenient assumption to
simplify the analytical solution of the steady state heat equation. In reality, the signal intensity in a single-mode fiber core 
follows a nearly Gaussian profile, so the piece-wise constant assumption is not strictly true in the core of the fiber. However,
it is important to asses the accuracy of this assumption, because the heating profile in the core is the main underlying  
factor in transitioning from the single-mode to multimode operation due to the thermo-optic effect.

In the following analysis, we borrow from the formalism presented by Bowman in Ref.~\cite{bowman2015}, primarily developed to analyze the
heat generation in low quantum defect lasers. The interested reader may consult that paper for further details. 

The total heat source density $q(x,y,z)$, i.e. the thermal energy deposited per second at the location $(x,y,z)$ inside the fiber in units of ${\rm W/m}^3$ is 
given by $q=q_{qd}+q_{fl}+q_{ap}+q_{as}$:
\begin{subequations}
\begin{align}
q_{qd}&=\left(\dfrac{N_t h c}{\tau_r}\right)
\dfrac{\lambda_s \beta_p i_p + \lambda_p \beta_s i_s + (\lambda_s - \lambda_p) (\beta_p - \beta_s) i_p i_s}{\lambda_s\lambda_p(1+i_p+i_s)},\\
q_{fl}&=-\left(\dfrac{N_t h c}{\tau_r}\right)\dfrac{\beta_p i_p +\beta_s i_s}{\lambda_f(1+i_p+i_s)},\\
q_{ap}&=\alpha_p I_p,\qquad q_{as}=\alpha_s I_s.
\end{align}
\end{subequations}
$q_{qd}$ is the contribution due to the quantum defect and $q_{fl}$  is the contribution from fluorescence where the negative sign indicates that it 
is a heat drain. $q_{ap}$ and $q_{as}$ are contributions due to the parasitic absorption of the pump and the signal, respectively.
$N_t$ is total Yb ion dopant density, $h$ is the Planck's constant, $c$ is the speed of light, and $\tau_r$ is the upper-level lifetime of Yb ions
doped in silica, which can be almost equal to its total lifetime~\cite{Mobini2019}. $\sigma^a_p$ and $\sigma^a_s$ are the absorption cross sections of the 
pump and signal, while $\sigma^e_p$ and $\sigma^e_s$ are the emission cross sections of the pump and signal, respectively. We also define:
\begin{align}
\beta_p=\dfrac{\sigma^a_p}{\sigma^a_p+\sigma^e_p},\quad \beta_s=\dfrac{\sigma^a_s}{\sigma^a_s+\sigma^e_s}.
\end{align}
$I_p$ and $I_s$ are the local pump and signal intensities, while $i_p$ and $i_s$ are the same quantities normalized by their corresponding saturation values. We have
\begin{subequations}
\begin{align}
i_p=\dfrac{I_p}{I_p^{\rm sat}},\qquad I_p^{\rm sat}=\dfrac{hc\beta_p}{\lambda_p\tau_r\sigma^a_p},\\
i_s=\dfrac{I_s}{I_s^{\rm sat}},\qquad I_s^{\rm sat}=\dfrac{hc\beta_s}{\lambda_s\tau_r\sigma^a_s}.
\end{align}
\end{subequations}
$\lambda_f$ is the mean fluorescence wavelength and is defined in Ref.~\cite{bowman2015}.

Unless the fiber laser or amplifier system is especially tuned to operate in a nearly radiation-balanced 
regime~\cite{bowman2010minimizing,bowman2016low,mobini2018thermal,mobiniCoreCladding}, the contribution from 
$q_{fl}$ can be ignored. In a conventional DCF laser or amplifier where the signal and
pump powers are nearly of the same order of magnitude, the signal intensity is substantially higher than the 
pump intensity because the signal propagates in a much smaller area in the core. It can be readily seen that 
the limit of $1\ll i_p\ll i_s$ commonly applies. In this limit, we have
\begin{align}
\label{Eq:tophat}
q_{qd}\approx \left(\dfrac{N_t h c}{\tau_r}\right) \dfrac{(\lambda_s - \lambda_p) \beta_p i_p}{\lambda_s\lambda_p}=N_t\sigma^a_pI_p\left(\dfrac{\lambda_s - \lambda_p}{\lambda_s}\right),
\end{align}
which is the formula that was presented earlier in section~\ref{sec:examples}\ref{sec:gamma}. Here, we have taken into account that $\beta_p\gg \beta_s$.
Equation~\ref{Eq:tophat} clearly shows that the heat due the quantum defect follows a top-hat profile in high-power DCF lasers and amplifiers, because it is proportional 
to $i_p$ rather than $i_s$ in the limit of $i_p\ll i_s$. Note that this top-hat form is enforced by the signal saturation effect in the core.
Also, in the inner-cladding where $i_s=0$, the heat density is also uniform as assumed in this paper.

Near the tail of the signal at the core-inner-cladding boundary, $i_p\ll i_s$ may no longer be valid, so that 
will present a deviation form the top-hat assumption for the heat generation. Also, if the input signal power 
is low in an amplifier set-up, $i_p\ll i_s$ may not apply near the input and the heat profile in the core may shape 
somewhere between a Gaussian and a top-hat depending on the specifics of the problem. 
Moreover, unlike $q_{ap}$ which is of the top-hat form, $q_{as}$ follows the near-Gaussian profile of the signal. If the contribution of $q_{as}$ is considerable,
the top-hat assumption must be revisited. 
The contribution from $q_{fl}$ must also be included in radiation-balanced lasers and amplifiers~\cite{bowman2010minimizing,bowman2016low,mobini2018thermal,mobiniCoreCladding}.
However, in all these cases, the piece-wise constant assumption for $q(\rho)$ in the radial coordinate should give a reasonably accurate assessment of the temperature profile
in the core if all the sources of heating are considered. Of course, the cladding temperature profiles are not affected and the piece-wise assumption for the
inner-cladding always holds. 
\section*{Appendix C: Most useful formulas}
In this Appendix, we summarize the equations that are most useful for direct comparison with experiments. In this Appendix, we only consider 
the case where the heat density due to quantum defect in the core ($Q_{qd}$) overwhelms other sources of heat, as in the case in conventional fiber lasers and amplifiers. The more general case 
is treated in detail in the main text of the paper. The most relevant equations are: 
\begin{subequations}
\begin{align}
\Delta T&=\dfrac{Q_{qd}}{2\pi c\,H},\\
\delta T_a&=\dfrac{Q_{qd}}{4\pi\kappa_a},\\
\delta T_b&=\ln\left(\dfrac{b^2}{a^2}\right)\dfrac{Q_{qd}}{4\pi\kappa_b},\\
\delta T_c&=\ln\left(\dfrac{c^2}{b^2}\right)\dfrac{Q_{qd}}{4\pi\kappa_c}.
\end{align}
\end{subequations}

For example, for $Q_{qd}=40\,{\rm W/m}$, outer diameter of $2c=245\,\mu{\rm m}$, core and inner-cladding of glass with 
$\kappa_a=\kappa_b=1.38\,{\rm W/(m.K)}$, outer-cladding of polymer with $\kappa_c=0.276\,{\rm W/(m.K)}$, and $H\approx 920\,{\rm W/(m^{2}.K)}$
for a moderate-flow-speed-water-cooled fiber, we obtain:
\begin{align*}
\Delta T=56^{\circ}{\rm C},\ \delta T_a=3.46^{\circ}{\rm C},\ \delta T_b=11.7^{\circ}{\rm C},\ \delta T_c=15.5^{\circ}{\rm C}.
\end{align*}
\section*{Acknowledgment}
The author is grateful to Dr. Thomas Schreiber of Fraunhofer Institute for Applied Optics and Precision Engineering, Jena, Germany, for helpful suggestions. 
\section*{Funding Information}
This material is based upon work supported by the Air Force Office of Scientific Research under award number FA9550-16-1-0362 
titled Multidisciplinary Approaches to Radiation Balanced Lasers (MARBLE).
\section*{Disclosures}
The author declares no conflicts of interest.
\bigskip

\providecommand{\noopsort}[1]{}\providecommand{\singleletter}[1]{#1}%


\end{document}